\documentclass[amsmath, amsfonts, superscriptaddress, twocolumn, prl]{revtex4}
\usepackage{graphicx}
\usepackage{epsfig}
\usepackage{bm}
\usepackage{dcolumn}
\usepackage{amsmath}
\usepackage{amssymb}
\usepackage{color}
\usepackage[varg]{txfonts}

\newcommand{\beg}{\begin{equation}}
\newcommand{\en}{\end{equation}}
\newcommand{\bp}{\mathbf p}

\newcommand{\br}{\mathbf r}

\newcommand{\up}{\uparrow}
\newcommand{\dn}{\downarrow}
\newcommand{\dg}{^\dagger}
\newcommand{\Tr}{\mathop{\rm Tr}\nolimits}

\begin{document}

\title{Josephsonic diagnostic of competing orders in quantum critical multiband superconductors}

\author{Maxim Dzero}
\affiliation{Department of Physics, Kent State University, Kent, Ohio 44242, USA}

\author{Alex Levchenko}
\affiliation{Department of Physics, University of Wisconsin-Madison, Madison, Wisconsin 53706, USA}

\date{May 12, 2018}

\begin{abstract}
Motivated by the recent experimentally observed manifestations of the quantum critical point fluctuations in the thermodynamic properties of multiband superconductors, we derive a general expression for the Josephson current of various junctions between two superconductors in the phase of superconductivity coexistence with the spin-density-wave. We demonstrate that the critical current peaks at the quantum critical point that separates pure and mixed superconducting phases. We argue that our results are generic and, in particular, can be adopted to explain the recent observations of a nonmonotonic dependence of the supercurrent on the external pressure in the heavy fermion superconductor CeRhIn$_5$, and on the chemical doping in iron-based superconductors such as Ba(Fe$_{1-x}$Co$_x$)$_2$As$_2$.
\end{abstract}

\maketitle

\section{I. Introduction and motivation} 

Competition of electronic pairing correlations in multiband metals leads to a formation of different long-range orderings most prominently in the form of superconductivity, itinerant magnetism, and various possible forms of density-waves. In some instances these collective electronic phases may coexist over the extended parts of the phase diagram. In particular, despite being different at the microscopic level, cuprates, iron-pnictides, and heavy-fermion compounds exhibit qualitatively similar global features at the level of phase diagram with dome-shaped superconductivity as a function of an experimental control parameter such as chemical doping or pressure (see reviews \cite{Norman,Pfleiderer,Chubukov} and references therein). In the context of iron-pnictides specifically, spin-density-wave (SDW) can coexist with superconductivity (SC) \cite{Fernandes-PD,Vorontsov-PD}. If line of the SDW transition extends all the way to zero temperature beneath the line of superconducting state it could ultimately terminate at the quantum critical point (QCP) \cite{Shibauchi}. This scenario is somewhat complicated by the fact that SDW occurs in proximity to a structural transition that also enters the superconducting dome. The possibility of having multiple quantum critical points surrounded by superconductivity was recently investigated theoretically \cite{Fernandes-QCPs}.

A magnetic QCP without superconductivity has been extensively studied \cite{Abanov,Metlitskii,Efetov}. Quantum fluctuations near this point are known to give rise to non-Fermi liquid behavior and to singularities in various electronic characteristics. SDW instability inside the $d$-wave SC state has been analyzed in Ref. \cite{Pelissetto} and was shown to lead to non-Fermi liquid behavior of nodal fermions. The observation of SC-SDW coexistence in iron-pnictides brought up the new issue of whether there are electronic singularities at a magnetic QCP which develops in the presence of an $s^{\pm}$-wave order. This problem attracted a great deal of interest recently. Measurements of the London penetration depth in BaFe$_2$(As$_{1-x}$P$_x$)$_2$ \cite{Hashimoto,Auslaender-1} and (Ba$_{1-x}$K$_x$)Fe$_2$As$_2$ \cite{Auslaender-2} revealed a sharp peak-feature near the putative QCP at the optimal doping leading a flurry of theoretical proposals to explain this effect \cite{Levchenko,Chowdhury-1,Nomoto,Chowdhury-2}.  An additional evidence was also obtained from the subsequent specific-heat-jump and de Haas-van Alphen effect experiments that point towards strong increase in the quasiparticle mass at the QCP \cite{Carrington-1}. The lower and upper superconducting critical fields were also found to behave anomalously \cite{Carrington-2}.  

Since gapped SC state expels quasiparticle degrees of freedom and makes it difficult to probe quantum critical behavior of electronic response functions, the question arises of whether a native observable to superconducting state, such as supercurrent, can provide an effective diagnostic for competing orders under superconducting dome. In this paper we give an affirmative answer to that question. One of our main results is that maximal current of the Josephson weak link between two pnictide superconductors peaks at the QCP of optimal-doping and a similar peak occurs in the thermodynamic critical current of a bulk sample. It is important to mention here that the early works on Josephson currents in iron-pnictides focused primarily on the issues of phase-sensitive probe that could reveal the pairing symmetry of the underlying order parameter and elucidate the nature of the induced anomalous proximity effect \cite{Mazin,Wu,Nagaosa,Linder,Tsai,Chen,Ota-1,Ota-2,Yerin,Berg,Golubov,Koshelev,Stanev,Lin,Vakaryuk,Apostolov}. The emphasis of our work is on the coexistence region, for which we derive several universal Josephson current-phase relationships for contacts of different types.    

\section{II. Microscopic model and phase diagram} 

\begin{figure}
\includegraphics[width=0.9\linewidth]{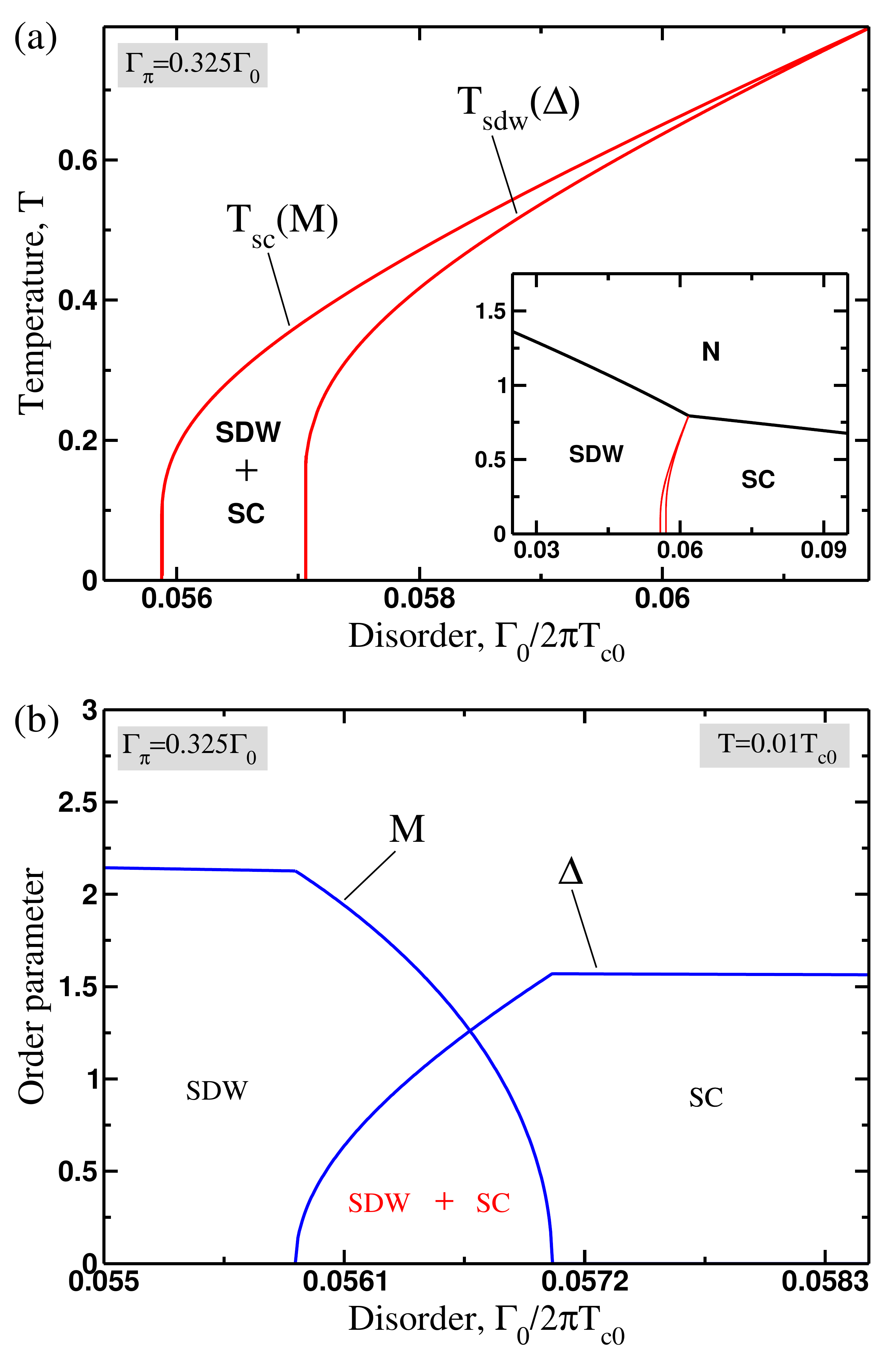} 
\caption{[Color online] Panel (a): The mean-field phase diagram for the two-band model as a function of intraband and interband disorder with an emphasis on the coexistence region. Inset shows the full expanded phase diagram on a larger scale of doping. Panel (b): The dependence of the superconducting and SDW order parameters on the intraband disorder scattering rate.}
\label{Fig-PD}
\end{figure}

To study the essential physics of the Josephsonic probe in the context of competing orders it will be sufficient to adopt the simplest minimal two-band model for the interplay between itinerant SDW and $s^{\pm}$-wave SC \cite{Fernandes-PD,Vorontsov-PD}. For that purpose, we consider a circular hole pocket at the center of the Brillouin zone and an electron pocket displaced from the center by $\bm{Q}=(0,\pi)$ [or $(\pi,0)$]. We assume that chemical doping acts as a source of disorder and induces both intra- and interband scattering. In an effective low-energy theory we account for particle-hole and particle-particle interaction channels with angle-independent couplings $\lambda_{\rm{sdw}}$ and $\lambda_{\rm{sc}}$, and high-energy cutoff $\Lambda$. We treat these interactions within a mean-field approximation, by introducing SC and SDW order parameters, $\Delta$ and $\bm{M}$, respectively, and decomposing the four-fermion interactions into effective quadratic terms via Hubbard-Stratonvich transformation. For this model we derive Eilenberger equation for the semiclassical Green function-$\hat{G}$ which reads \cite{Vavilov,Moor,Dzero}
\begin{equation}\label{G}
[i\omega_n\hat{\tau}_3\hat{\rho}_3\hat{\sigma}_0,\hat{G}]-\left[\hat{H}\hat{\tau}_3\hat{\rho}_3\hat{\sigma}_0,\hat{G}]-[\hat{\Sigma}\hat{\tau}_3\hat{\rho}_3\hat{\sigma}_0,\hat{G}\right]=i{\bm v}_{F}\cdot{\mbox{\boldmath $\nabla$}} \hat{G}
\end{equation}  
where $\omega_n$ is the Matsubara frequency, $\bm{v}_F$ is the Fermi velocity, and brackets stand for commutators of matrices. The interaction part of the mean-field Hamiltonian has the form 
\begin{equation}\label{H}
\hat{H}=-\Delta\hat{\tau}_3\hat{\rho}_2\hat{\sigma}_2+M\hat{\tau}_1\hat{\rho}_3\hat{\sigma}_3,
\end{equation}
which is written in the basis of two flavor of fermions 
\beg\label{PsiBasis}
\overline{\Psi}_\bp=\left(\hat{c}_{\bp\up}\dg, \hat{c}_{\bp\dn}\dg, \hat{c}_{-\bp\up}, \hat{c}_{-\bp\dn},
\hat{f}_{\bp\up}\dg, \hat{f}_{\bp\dn}\dg, \hat{f}_{-\bp\up}, \hat{f}_{-\bp\dn}\right)
\en
that correspond to electron and hole pockets of the two-band model. Three sets of Pauli matrices $\{\hat{\rho}_i,\hat{\sigma}_i,\hat{\tau}_i\}$ are needed to describe Gorkov-Nambu, spin, and band sub-spaces. The expression for the disorder potential in this basis reads
\begin{equation}\label{disorder}
\hat{U}(\bm{r})=\sum\limits_{l}\left[U_0\hat{\tau}_0\hat{\rho}_3\hat{\sigma}_0+U_\pi \hat{\tau}_1\hat{\rho}_3\hat{\sigma}_0\right]\delta(\br-{\mathbf R}_l),
\end{equation} 
where summation goes over the impurities at random locations ${\mathbf R}_l$. The first term in this expression describes the scattering within each band, while the second term scatters quasiparticles between the two bands. The self-energy due to the scattering of the quasiparticles on disorder potential can be found by the standard methods of the many-body theory. At the level of the self-consistent Born approximation we find the following self-energy 
\begin{align}\label{Sigma}
\hat{\Sigma}=-i\Gamma_0\hat{\tau}_3\hat{\rho}_0\hat{\sigma}_0\int\frac{d\Omega}{4\pi}\hat{G}(i\omega_n,{\bm v}_F,{\bm r})\hat{\tau}_0\hat{\rho}_3\hat{\sigma}_0\nonumber\\ -i\Gamma_\pi\hat{\tau}_2\hat{\rho}_0\hat{\sigma}_0\int\frac{d\Omega}{4\pi}\hat{G}(i\omega_n,{\bm v}_F,{\bm r})\hat{\tau}_1\hat{\rho}_3\hat{\sigma}_0,
\end{align} 
where $\Gamma_0= \pi\nu_Fn_{\rm{imp}}|U_0|^2/4$ and $\Gamma_\pi=\pi\nu_Fn_{\rm{imp}}|U_\pi|^2/4$ are the intraband and interband scattering rates that are proportional to the impurity concentration $n_{\rm{imp}}$, the total quasiparticle density of states at the Fermi energy $\nu_F$, and strength of the disorder potential $U$. The integration in Eq. \eqref{Sigma} is performed over the directions of the Fermi velocity. Without loss of generality, we assumed in Eq. \eqref{H} $\bm{M}$ is along $z$-axis. Finally, SC and SDW order parameters are subject to the self-consistency equations 
\begin{equation}\label{Delta-M}
\begin{split}
\frac{\Delta}{\nu_F\lambda_{\rm{sc}}}&=2\pi T\sum^\Lambda_{\omega_n}\Tr\left[\hat{G}(\hat{\tau}_0+\hat{\tau}_3)\hat{\rho}^{+}\hat{\sigma}^{+}\right],\\
\frac{M}{\nu_F\lambda_{\rm{sdw}}}&=2\pi T\sum^\Lambda_{\omega_n}\Tr\left[\hat{G}\hat{\tau}^+(\hat{\rho}_0+\hat{\rho}_3)\hat{\sigma}_3\right],
\end{split}
\end{equation} 
where we adopted the following notation \linebreak $\hat{\sigma}^{+}=(\hat{\sigma}_1+i\hat{\sigma}_2)/2$. By combining Eqs. (\ref{G}) and \eqref{Delta-M} under assumption that all quantities remain spatially homogeneous we reconstruct the temperature-intraband scattering rate $(T,\Gamma_0)$ phase diagram shown in Fig. \ref{Fig-PD}(a) for a particular value of interband scattering $\Gamma_\pi$. The corresponding self-consistent order parameters are shown on Fig. \ref{Fig-PD}(b). The scale of plots is normalized to $T_{c0}=1.13\Lambda\exp[-2/(\nu_F\lambda_{\rm{sc}})]$, which is the bare critical temperature of SC state in the absence of disorder and SDW. The $T_{\rm{sc}}(M)$ line defines the part of superconducting dome that emerges from the preexisting SDW order. The $T_{\rm{sdw}}(\Delta)$ line separates mixed and pure phases, and terminates at the QCP for an optimal doping level of $\Gamma_0/(2\pi T_{c0})\approx 0.057$. This model was successfully used in the past to describe the main qualitative features of the doping-dependence for the specific heat jump \cite{Vavilov}, Knight shift \cite{Moor}, enhancement of superconducting critical temperature by disorder \cite{Fernandes-Tc}, and explain nonmonotonic behavior of the London penetration depth \cite{Dzero}. The only drawback of the model is that it yields a very narrow region of coexistences between $\Delta$ and $M$ in the parameter space set by $\Gamma_{0,\pi}$, yet it captures the main physics and is appealing because of its simplicity.  

\begin{figure*}
\includegraphics[width=0.3\linewidth]{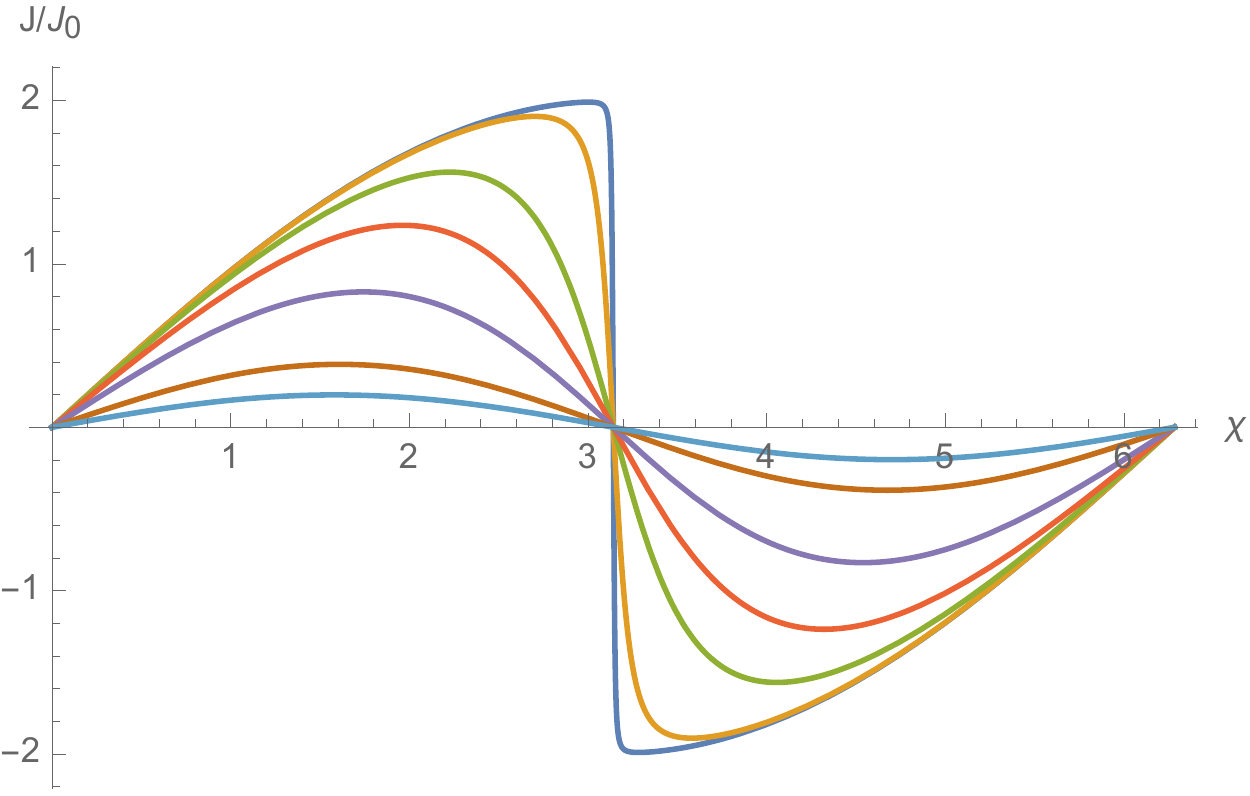}
\includegraphics[width=0.3\linewidth]{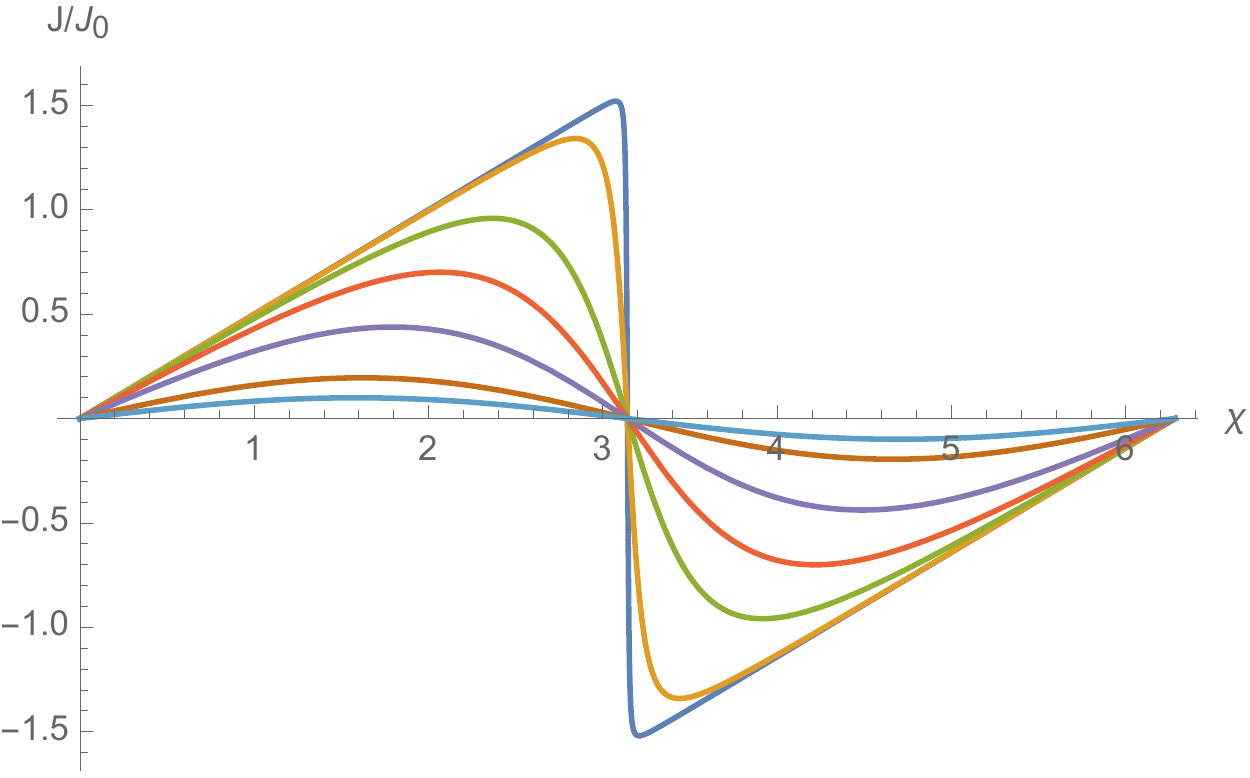}
\includegraphics[width=0.3\linewidth]{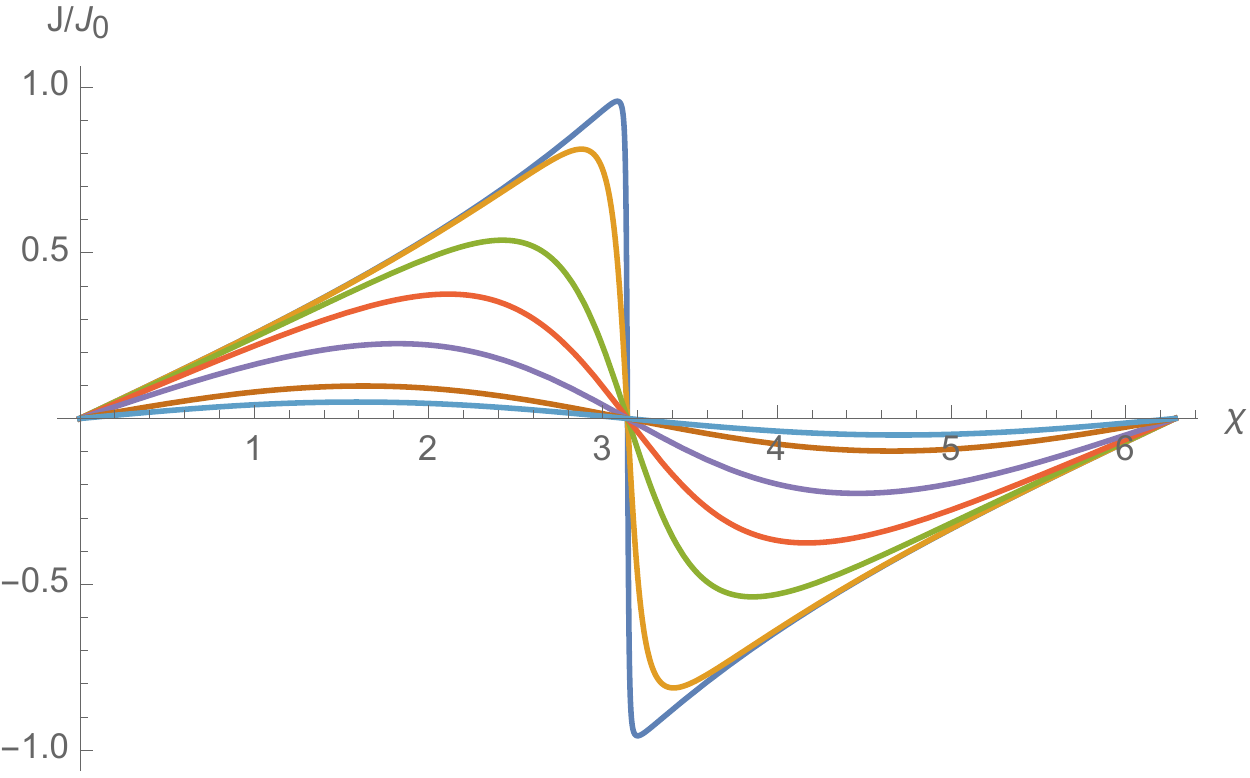}
\caption{Josephson current-phase relationship for different $m=\{0.005,0.05,0.25,0.5,1,2.5,5\}$ plotted from Eq. \eqref{J-CPR}. Current is normalized in units $J_0=c_\alpha\Delta/(eR_N)$. Left panel corresponds to $\alpha=3/2$, middle to $\alpha=1$, while the right panel corresponds to $\alpha=1/2$}  
\label{Fig-J-CPR}	
\end{figure*}

\section{III. Josephson current-phase-relations and critical current}

In what follows, we envision a Josephson junction made of two FeSCs. We assume that the length $L$ of the weak link between them is small compared to both superconducting and magnetic coherent lengths $L\ll\{v_F/\Delta,v_F/M\}$. This limit enables us to treat the junction as an effective multi-channel point contact with transmission eigenmodes labeled by $D\in [0,1]$. We consider a quasi-one-dimensional geometry where all functions depend on a single coordinate-$x$ which is a direction along the junction with the contact located at $x=0$. Lastly, it is important to keep in mind that within the semiclassical formalism Eilenberger Green function must satisfy a nonlinear boundary condition \cite{Zaitsev1984,Yip1997}:
\begin{equation}\label{B}
\hat{G}_a(0)\left(R\hat{G}^2_{s+}(0)+\hat{G}^2_{s-}(0)\right)=D\hat{G}_{s-}(0)\hat{G}_{s+}(0)
\end{equation}
where $\hat{G}_{s(a)}=[\hat{G}(i\omega_n,v_x,x)\pm\hat{G}(i\omega_n,-v_x,x)]/2$, $\hat{G}_{s\pm}(0)=[\hat{G}_{s}(+0)\pm G_{s}(-0)]/2$, and $R=1-D$ is the reflection coefficient from the interface. Provided that one knows the solution to Eq. \eqref{G} and able to resolve the nonlinear matrix constraint in Eq. \eqref{B}, the current across the junction can be expressed in terms of the Eilenberger function as follows
\begin{equation}\label{J}
J=e\nu_Fv_FT\!\sum_{\omega_n}\!\!\int^{\frac{\pi}{2}}_{0}\!\!\mathrm{Im}\left\{\Tr
[\hat{\tau}_0\hat{\rho}_3\hat{\sigma}_0\hat{G}_a(i\omega_n,v_F\sin\phi,0)]\right\}\sin\phi\frac{d\phi}{2\pi}
\end{equation}     
In realistic contacts scattering is rather described by a continuous transmission distribution $\rho(D)$ than by discrete transmission eigenvalues $D$. It is thus of practical importance to find an averaged Josephson current. There are several generic contact types that have been discussed in the literature in the context of mesoscopic transport. Their distributions are described by the function of the form
\begin{equation}\label{DTE}
\rho(D)\propto \frac{1}{D^\alpha\sqrt{1-D}},\quad \alpha=1/2,1,3/2. 
\end{equation}   
The case with the power exponent $\alpha=1/2$ corresponds to two ballistic connectors with equal conductances in series \cite{Baranger}. The case with $\alpha=1$ corresponds to a diffusive connector \cite{Dorokhov}. The case with $\alpha=3/2$ describes an interface with a high density of randomly distributed scatterers \cite{Schep}. 

It is important to realize that Eq. \eqref{B} can be brought to the form analogous to the circuit-theory boundary conditions of Andreev refection \cite{Nazarov}. Indeed, it can be shown that the nonlinear boundary condition can be equivalently rewritten in terms of the bounded quasiclassical functions $\hat{G}^{r/l}_{\mathrm{b}}$ 
\begin{equation}\label{B-b}
\left(2-D+\frac{D}{2}\left\{\hat{G}^r_{\mathrm{b}}(0),\hat{G}^l_{\mathrm{b}}(0)\right\}\right)\hat{G}_a(0)=\frac{D}{2}\left[\hat{G}^r_{\mathrm{b}}(0),\hat{G}^l_{\mathrm{b}}(0)\right].
\end{equation}
These functions from the left/right ($l/r$) side of the junction can be computed separately by employing the method of finding the auxiliary solutions to the quasiclassical equations (see Ref. \cite{Yip1997} for the detailed description of this method). The calculation is tedious so we omit technical details and present only the final result 
\begin{equation}\label{G-b}
\hat{G}^{r/l}_{\mathrm{b}}(0)=-\frac{1}{E_{nj}}(\omega_n\hat{\tau}_3\hat{\rho}_3\hat{\sigma}_0+\Delta_j\hat{\tau}_0\hat{\rho}_1\hat{\sigma}_2+M_j\hat{\tau}_2\hat{\rho}_0\hat{\sigma}_3)
\end{equation}
where $\Delta_j$ and $M_j$ for $j=1,2$ are SC and SDW order parameters from the left/right side of the junction, and $E_{nj}=\sqrt{\omega^2_n+\Delta^2_j+M^2_j}$. We use this expression in Eq. \eqref{B-b}, compute commutators and take a trace to arrive at  
\begin{equation}\label{Tr}
\mathrm{Tr}[\hat{\tau}_0\hat{\rho}_3\hat{\sigma}_0\hat{G}_a]\!=\!
\frac{8iD\Delta_1\Delta_2\sin\chi}{(2-D)E_{n1}E_{n2}+D(\omega^2_n+M_1M_2+\Delta_1\Delta_2\cos\chi)}
\end{equation}
In this formula $\Delta_j$ should be understood as absolute values of the gaps whereas the global phase difference $\chi$ across the junction was accounted explicitly.  Equation \eqref{Tr} gives us Josephson current-phase relation as defined by Eq. \eqref{J} and the distribution function \eqref{DTE} allows us to find its average:
\begin{equation}\label{J-Average}
J(\chi)=\frac{2\pi T}{eR_N}\sum_{\omega_n}\frac{\Delta_1\Delta_2\sin\chi}{E_{n1}E_{n2}}
\int\frac{D\rho(D)dD}{(2-D)+Dp(\chi)},
\end{equation}
where $R_N$ is the normal state resistance of the junction, $\chi$ is the global phase difference of superconducting order parameters across the junction, and the phase-dependent parameter is 
\begin{equation}
p(\chi)=\frac{\omega^2_n+M_1M_2+\Delta_1\Delta_2\cos\chi}{\sqrt{\omega^2_n+M^2_1+\Delta^2_1}\sqrt{\omega^2_n+M^2_2+\Delta^2_2}}.
\end{equation}
Integration over $D$ can be completed in elementary functions for all the cases under consideration. At zero-temperature, the remaining Matsubara sum can be converted into an integral, $2\pi T\sum_{\omega_n}\to\int d\omega$, that also can be computed analytically in the case of a symmetric contact with identical order parameters on both sides of the junction. For this case specifically we can compactly write the current 
\begin{equation}\label{J-CPR}
J(\chi)=J_{0} f_\alpha(\chi,m)
\end{equation}
with $m=M/\Delta$, $J_{0}=c_\alpha{\Delta}/{eR_N}$, the numerical coefficients $c_\alpha$ for $\alpha=1/2,1,3/2$ are all of the order of unity, and $f_\alpha$ are universal dimensionless functions whose explicit expressions read
\begin{equation}
\begin{split}
&f_{3/2}=\frac{\sin(\chi)}{\sqrt{m^2+\cos^2(\chi/2)}}, \\ &f_1=\cos(\chi/2)\sqrt{\frac{m^2+1}{m^2+\cos^2(\chi/2)}}\arctan\left(\frac{\sin(\chi/2)}{m^2+\cos^2(\chi/2)}\right), \\
&f_{1/2}=\cot(\chi/2)\sqrt{m^2+1}\left(\sqrt{\frac{m^2+1}{m^2+\cos^2(\chi/2)}}-1\right).
\end{split}
\end{equation}

\begin{figure}
\includegraphics[width=0.9\linewidth]{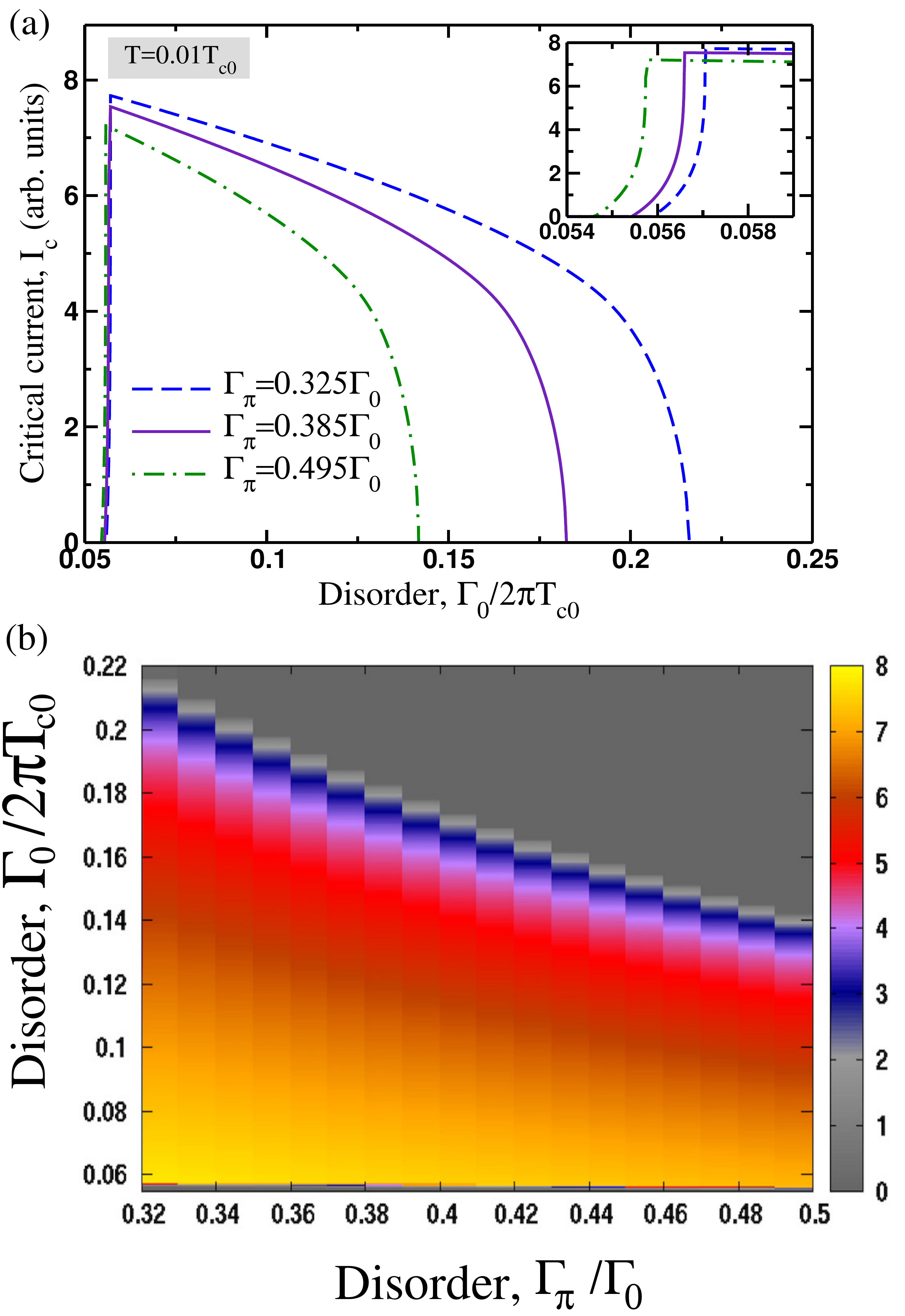} 
\caption{[Color online] Panel (a): critical current as a function of the intraband scattering rate for a fixed ratio $\Gamma_\pi/\Gamma_0$. Panel (b): contour plot of the critical current (arb. units) as a function of $\Gamma_{\pi}/\Gamma_0$ and $\Gamma_0$. }
\label{Fig-Ic}
\end{figure}

The corresponding current-phase relationships are plotted in Fig. \ref{Fig-J-CPR}. They display strong sensitivity to magnetic order parameter and anharmonic non-sinusoidal behavior. The two limiting case of interest are weak and strong superconductivity that correspond to looking at different parts of the phase diagram [see Fig. \ref{Fig-PD}(ab)]. In the case when $M\gg\Delta$, which occurs close to $T_{\mathrm{sc}}(M)$ line of the phase diagram, maximal Josephson current is suppressed, $J_c\propto \Delta^2/(eR_NM)$, for all contact types discussed above. In the opposite limit, $\Delta\gg M$, that is relevant in the proximity to $T_{\mathrm{sdw}}(\Delta)$ line which separates mixed and pure phase, Josephson current amplitude saturates to a maximum set by superconducting gap, $J_c\propto \Delta/eR_N$. 

Our model allows to study doping evolution of the current across the entire range of parameters that define the phase diagram. The results are shown on Fig. \ref{Fig-Ic}. The initial sharp increase of the current on panel-(a) corresponds to the regime when magnetic order is suppressed quicker by disorder than superconducting order. This happens because $M$ is sensitive to the total scattering rate $\Gamma_0+\Gamma_\pi$, whereas superconducting order is suppressed by interband scattering only. Such a steep dependence is, however, an artifact of the model where coexistence region is very narrow. Nevertheless it captures the main effect that current peaks at the optimal doping that sets the critical point beneath the dome. In the overdoped regime current amplitude gradually decays as superconducting order parameter bends to zero [see Fig. \eqref{Fig-PD}], which happens progressively quicker for higher values of interband scattering $\Gamma_\pi$ that acts as an effective pair-breaking for $s^{\pm}$ superconducting state. 

\begin{figure}
\includegraphics[width=0.9\linewidth]{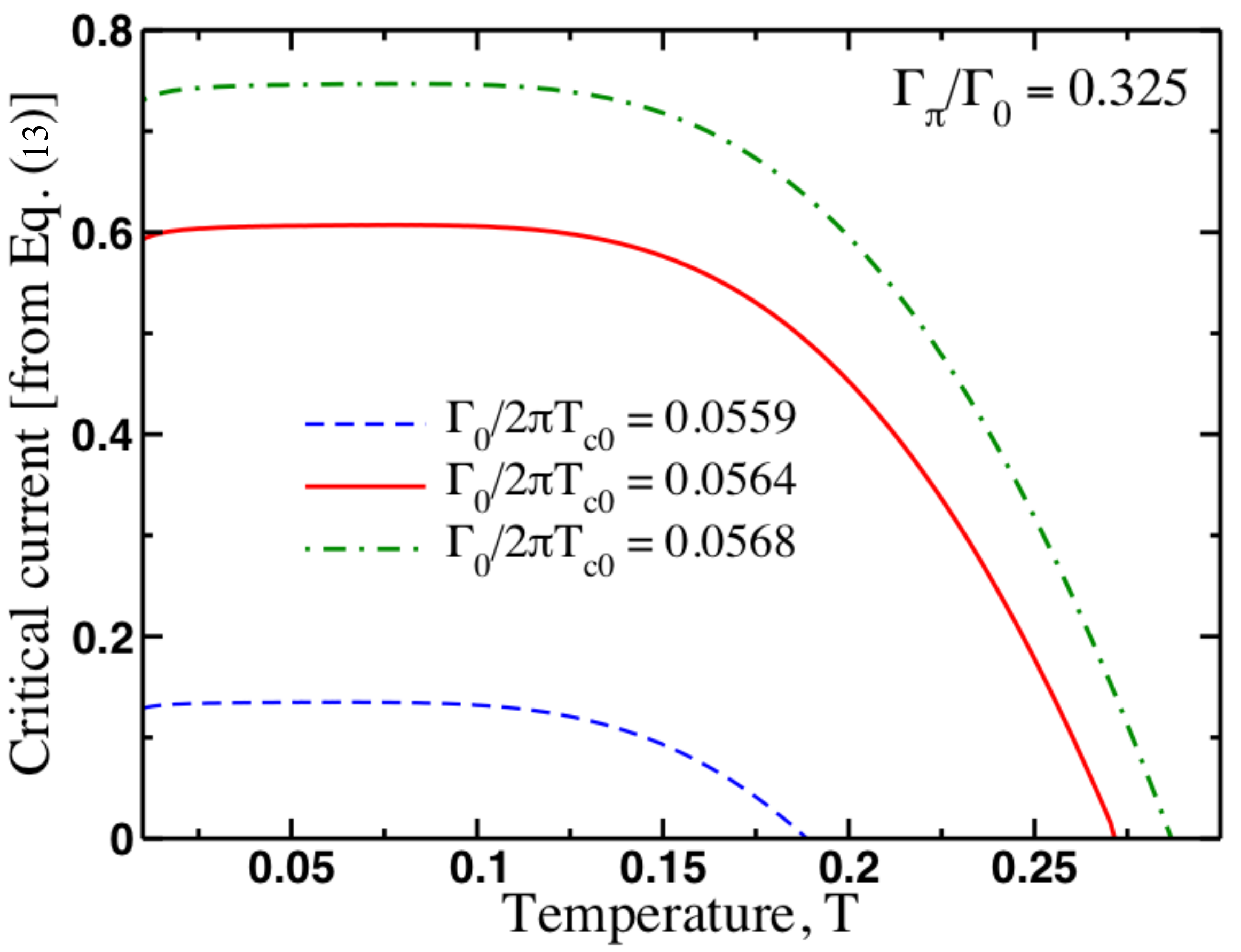} 
\caption{[Color online] Temperature dependence for the critical current in the model of the contact when $\alpha=3/2$. The scale of temperature is normalized to $T_{c0}$ and scattering parameters $\Gamma_{0,\pi}$ for all three curves are chosen to lie inside the coexistence region.}
\label{Fig-Ic-T}
\end{figure}

The general formula in Eq. \eqref{J-Average} also enables us to study the temperature dependence of the critical current. This computation has to be done numerically as temperature enters directly in the Matsubara sum, and also via the order parameters as determined by self-consistency equations. From the plots presented in Fig. \ref{Fig-Ic-T} it becomes clear that current is primarily determined by the temperature dependence of the gap which manifests by a rapid saturation of the current in the low temperature range.   

\section{IV. Discussions and outlook} 

We presented results for the critical Josephson current and current-phase relationship of pnictide superconductor junctions with the focus on the coexistence region of superconducting and spin-density wave order. We studied several types of junctions and used generic circuit-theory boundary conditions within the underlying semiclassical Eilenberger approach. Our main result concerns the doping dependence of the maximal attainable current that exhibits a peak that is correlated with the location of the quantum critical point hidden under the dome of superconductivity. This makes Josephsonic response to be an effective diagnostic of competing orders in addition to experimental techniques that exploit measurements of the magnetic penetration depth at low temperatures. 

A few comments are in order in relation to the results we presented in this paper. We performed calculations at the mean field level and the peak of the critical current has a cusp dependence at it maximum [see Fig. \ref{Fig-Ic}]. The account for magnetic quantum fluctuations near the critical point will likely smear this feature but it remains to be verified with the explicit computation. In addition, this model can be generalized to include nodes of the superconducting order parameter. This effect will be important to understand how critical current scales with temperature and along $T_{\mathrm{sdw}}(\Delta)$ second-order phase transition boundary in particular. We expect that peak of $J_c$ is pinned to $T_{\rm{sdw}}(\Delta)$ line beneath the dome of superconductivity.  

We also comment on the relevance of our theory to recent experiments and directions for further developments. The results presented in this work are perhaps the most relevant in the context of thermodynamic $J_c$ measurements in Ba(Fe$_{1-x}$Co$_x$)$_2$As$_2$ \cite{Ishida}. It is known that cobalt substitution adds appreciable amount of scattering so the application of the model when doping acts as a source of disorder is natural. Experiment shows that magnitude of $J_c$ significantly changes with $x$, exhibiting a sharp maximum at $x\sim0.057$, which is a slightly underdoped composition for Co-Ba122 that marks the onset of the coexistence between the itinerant antiferromagnetic and superconducting phases.  This behavior is consistent with predictions of the model that we explored in this paper. In addition, it has been demonstrated, that under controlled electron irradiation that adds nonmagnetic point defects, the topology of the superconducting gap in BaFe$_2$(As$_{1-x}$P$_x$)$_2$ changes from being nodal to gapped and back to nodal but of the other symmetry \cite{Mizukami}. The results were interpreted from the analysis of the low-temperature scaling of the London penetration depth. Our theory can be fruitful to address and model this peculiar behavior. As Josephson effect provides a natural phase-sensitive probe it will be of interest to investigate behavior of the critical current for various possible symmetries of the order parameters and different contact types. 

Thermodynamic critical current has been recently measured in the bulk sample of CeRhIn$_5$ \cite{Bauer2018} and in films of YBa$_2$Cu$_3$O$_{7-\delta}$ as well as other high-$T_c$ superconductors \cite{Talantsev}. In the heavy-fermion case $J_c$ peaks at the critical pressure where antiferromagnetic line ends inside superconducting dome. In the cuprate case, $J_c$ peaks on a critical hole doping where the pseudogap boundary line projects to $T=0$. Obviously these critical currents differ from the maxim current of the weak link in a Josephson junction, yet we argue that all of them will exhibit the same feature of having a peak at the optimal value of the control parameter that separates SC and the mixed phase with other competing order. This can be understood in our model as well in the context of FeSCs. Indeed, for simplicity of argument consider a part of the phase diagram in the proximity to $T_{\mathrm{sc}}(M)$. Near that line SC order parameter is small, and Ginzburg-Landau expansion is justified. The difference to a conventional case is that quadratic and quartic coefficients of this expansion in powers of $\Delta$ will vary strongly with $M$ as SC emerges from preexisting magnetic state. Then one can consider a supercurrent carrying state and follow the textbook procedure \cite{TheoryOfMetals} to find critical current. The results will exhibit a sharp rise of $J_c$ when doping is varied towards the tetracritical point of the phase diagram since $M$ is diminished, followed by a decay when superconductivity is extinguished at higher dopings. We thus take a point of view that peak effect in the supercurrents near critical point is a generic property of a wide class of quantum critical superconductors.

As a concluding remark we wish to mention that it is of conceptual theoretical interest to study hybrid circuits of FeSCs with inhomogeneous doping were an optimally doped superconducting film of iron-pnictide is proximitized on the surface of a parent material that is in a magnetic SDW phase. In this multilayered structure a proximity induced superconductivity may develop in nonsuperconducting part of the device. Conversely, an inverse proximity effect is possible where SDW order may penetrate inside the superconducting part of the device. The system is rather unique, as the use of the same parent compound with different doping should form nearly perfect interface between superconducting and magnetic phases. The extended Eilenberger formalism developed in this paper is exactly tailored to tackle such problems and should generate further developments and applications.  

\subsection{Acknowledgments} 

We would like to thank A. Balatsky, B. Davidson, Y. Matsuda, M. Rzchowski, T. Shibauchi, and M. Vavilov for very useful discussions. M.D. acknowledges the financial support by the National Science Foundation under Grant NSF-DMR-1506547 and, in part, by the U.S. Department of Energy, Office of Science, Office of Basic Energy Sciences under Award No. DE-SC0016481. The work of A.L. was financially supported in part by NSF CAREER Grant No. DMR-1653661 and the Wisconsin Alumni Research Foundation. 
  


\begin{thebibliography}{51}
\expandafter\ifx\csname natexlab\endcsname\relax\def\natexlab#1{#1}\fi
\expandafter\ifx\csname bibnamefont\endcsname\relax
  \def\bibnamefont#1{#1}\fi
\expandafter\ifx\csname bibfnamefont\endcsname\relax
  \def\bibfnamefont#1{#1}\fi
\expandafter\ifx\csname citenamefont\endcsname\relax
  \def\citenamefont#1{#1}\fi
\expandafter\ifx\csname url\endcsname\relax
  \def\url#1{\texttt{#1}}\fi
\expandafter\ifx\csname urlprefix\endcsname\relax\def\urlprefix{URL }\fi
\providecommand{\bibinfo}[2]{#2}
\providecommand{\eprint}[2][]{\url{#2}}

\bibitem[{\citenamefont{Norman and P{\'e}pin}(2003)}]{Norman}
\bibinfo{author}{\bibfnamefont{M.~R.} \bibnamefont{Norman}} \bibnamefont{and}
  \bibinfo{author}{\bibfnamefont{C.}~\bibnamefont{P{\'e}pin}},
  \bibinfo{journal}{Reports on Progress in Physics}
  \textbf{\bibinfo{volume}{66}}, \bibinfo{pages}{1547} (\bibinfo{year}{2003}).

\bibitem[{\citenamefont{Pfleiderer}(2009)}]{Pfleiderer}
\bibinfo{author}{\bibfnamefont{C.}~\bibnamefont{Pfleiderer}},
  \bibinfo{journal}{Rev. Mod. Phys.} \textbf{\bibinfo{volume}{81}},
  \bibinfo{pages}{1551} (\bibinfo{year}{2009}).

\bibitem[{\citenamefont{Chubukov}(2012)}]{Chubukov}
\bibinfo{author}{\bibfnamefont{A.}~\bibnamefont{Chubukov}},
  \bibinfo{journal}{Annual Review of Condensed Matter Physics}
  \textbf{\bibinfo{volume}{3}}, \bibinfo{pages}{57} (\bibinfo{year}{2012}).

\bibitem[{\citenamefont{Fernandes and Schmalian}(2010)}]{Fernandes-PD}
\bibinfo{author}{\bibfnamefont{R.~M.} \bibnamefont{Fernandes}}
  \bibnamefont{and}
  \bibinfo{author}{\bibfnamefont{J.}~\bibnamefont{Schmalian}},
  \bibinfo{journal}{Phys. Rev. B} \textbf{\bibinfo{volume}{82}},
  \bibinfo{pages}{014521} (\bibinfo{year}{2010}).

\bibitem[{\citenamefont{Vorontsov et~al.}(2010)\citenamefont{Vorontsov,
  Vavilov, and Chubukov}}]{Vorontsov-PD}
\bibinfo{author}{\bibfnamefont{A.~B.} \bibnamefont{Vorontsov}},
  \bibinfo{author}{\bibfnamefont{M.~G.} \bibnamefont{Vavilov}},
  \bibnamefont{and} \bibinfo{author}{\bibfnamefont{A.~V.}
  \bibnamefont{Chubukov}}, \bibinfo{journal}{Phys. Rev. B}
  \textbf{\bibinfo{volume}{81}}, \bibinfo{pages}{174538}
  (\bibinfo{year}{2010}).

\bibitem[{\citenamefont{Shibauchi et~al.}(2014)\citenamefont{Shibauchi,
  Carrington, and Matsuda}}]{Shibauchi}
\bibinfo{author}{\bibfnamefont{T.}~\bibnamefont{Shibauchi}},
  \bibinfo{author}{\bibfnamefont{A.}~\bibnamefont{Carrington}},
  \bibnamefont{and} \bibinfo{author}{\bibfnamefont{Y.}~\bibnamefont{Matsuda}},
  \bibinfo{journal}{Annual Review of Condensed Matter Physics}
  \textbf{\bibinfo{volume}{5}}, \bibinfo{pages}{113} (\bibinfo{year}{2014}).

\bibitem[{\citenamefont{Fernandes et~al.}(2013)\citenamefont{Fernandes, Maiti,
  W\"olfle, and Chubukov}}]{Fernandes-QCPs}
\bibinfo{author}{\bibfnamefont{R.~M.} \bibnamefont{Fernandes}},
  \bibinfo{author}{\bibfnamefont{S.}~\bibnamefont{Maiti}},
  \bibinfo{author}{\bibfnamefont{P.}~\bibnamefont{W\"olfle}}, \bibnamefont{and}
  \bibinfo{author}{\bibfnamefont{A.~V.} \bibnamefont{Chubukov}},
  \bibinfo{journal}{Phys. Rev. Lett.} \textbf{\bibinfo{volume}{111}},
  \bibinfo{pages}{057001} (\bibinfo{year}{2013}).

\bibitem[{\citenamefont{Abanov et~al.}(2003)\citenamefont{Abanov, Chubukov, and
  Schmalian}}]{Abanov}
\bibinfo{author}{\bibfnamefont{A.}~\bibnamefont{Abanov}},
  \bibinfo{author}{\bibfnamefont{A.~V.} \bibnamefont{Chubukov}},
  \bibnamefont{and}
  \bibinfo{author}{\bibfnamefont{J.}~\bibnamefont{Schmalian}},
  \bibinfo{journal}{Advances in Physics} \textbf{\bibinfo{volume}{52}},
  \bibinfo{pages}{119} (\bibinfo{year}{2003}).

\bibitem[{\citenamefont{Metlitski and Sachdev}(2010)}]{Metlitskii}
\bibinfo{author}{\bibfnamefont{M.~A.} \bibnamefont{Metlitski}}
  \bibnamefont{and} \bibinfo{author}{\bibfnamefont{S.}~\bibnamefont{Sachdev}},
  \bibinfo{journal}{Phys. Rev. B} \textbf{\bibinfo{volume}{82}},
  \bibinfo{pages}{075128} (\bibinfo{year}{2010}).

\bibitem[{\citenamefont{Efetov et~al.}(2013)\citenamefont{Efetov, Meier, and
  P{\'e}pin}}]{Efetov}
\bibinfo{author}{\bibfnamefont{K.~B.} \bibnamefont{Efetov}},
  \bibinfo{author}{\bibfnamefont{H.}~\bibnamefont{Meier}}, \bibnamefont{and}
  \bibinfo{author}{\bibfnamefont{C.}~\bibnamefont{P{\'e}pin}},
  \bibinfo{journal}{Nature Physics} \textbf{\bibinfo{volume}{9}},
  \bibinfo{pages}{442 EP } (\bibinfo{year}{2013}).

\bibitem[{\citenamefont{Pelissetto et~al.}(2008)\citenamefont{Pelissetto,
  Sachdev, and Vicari}}]{Pelissetto}
\bibinfo{author}{\bibfnamefont{A.}~\bibnamefont{Pelissetto}},
  \bibinfo{author}{\bibfnamefont{S.}~\bibnamefont{Sachdev}}, \bibnamefont{and}
  \bibinfo{author}{\bibfnamefont{E.}~\bibnamefont{Vicari}},
  \bibinfo{journal}{Phys. Rev. Lett.} \textbf{\bibinfo{volume}{101}},
  \bibinfo{pages}{027005} (\bibinfo{year}{2008}).

\bibitem[{\citenamefont{Hashimoto et~al.}(2012)\citenamefont{Hashimoto, Cho,
  Shibauchi, Kasahara, Mizukami, Katsumata, Tsuruhara, Terashima, Ikeda,
  Tanatar et~al.}}]{Hashimoto}
\bibinfo{author}{\bibfnamefont{K.}~\bibnamefont{Hashimoto}},
  \bibinfo{author}{\bibfnamefont{K.}~\bibnamefont{Cho}},
  \bibinfo{author}{\bibfnamefont{T.}~\bibnamefont{Shibauchi}},
  \bibinfo{author}{\bibfnamefont{S.}~\bibnamefont{Kasahara}},
  \bibinfo{author}{\bibfnamefont{Y.}~\bibnamefont{Mizukami}},
  \bibinfo{author}{\bibfnamefont{R.}~\bibnamefont{Katsumata}},
  \bibinfo{author}{\bibfnamefont{Y.}~\bibnamefont{Tsuruhara}},
  \bibinfo{author}{\bibfnamefont{T.}~\bibnamefont{Terashima}},
  \bibinfo{author}{\bibfnamefont{H.}~\bibnamefont{Ikeda}},
  \bibinfo{author}{\bibfnamefont{M.~A.} \bibnamefont{Tanatar}},
  \bibnamefont{et~al.}, \bibinfo{journal}{Science}
  \textbf{\bibinfo{volume}{336}}, \bibinfo{pages}{1554} (\bibinfo{year}{2012}),
  ISSN \bibinfo{issn}{0036-8075}.

\bibitem[{\citenamefont{Lamhot et~al.}(2015)\citenamefont{Lamhot, Yagil,
  Shapira, Kasahara, Watashige, Shibauchi, Matsuda, and
  Auslaender}}]{Auslaender-1}
\bibinfo{author}{\bibfnamefont{Y.}~\bibnamefont{Lamhot}},
  \bibinfo{author}{\bibfnamefont{A.}~\bibnamefont{Yagil}},
  \bibinfo{author}{\bibfnamefont{N.}~\bibnamefont{Shapira}},
  \bibinfo{author}{\bibfnamefont{S.}~\bibnamefont{Kasahara}},
  \bibinfo{author}{\bibfnamefont{T.}~\bibnamefont{Watashige}},
  \bibinfo{author}{\bibfnamefont{T.}~\bibnamefont{Shibauchi}},
  \bibinfo{author}{\bibfnamefont{Y.}~\bibnamefont{Matsuda}}, \bibnamefont{and}
  \bibinfo{author}{\bibfnamefont{O.~M.} \bibnamefont{Auslaender}},
  \bibinfo{journal}{Phys. Rev. B} \textbf{\bibinfo{volume}{91}},
  \bibinfo{pages}{060504} (\bibinfo{year}{2015}).

\bibitem[{\citenamefont{Almoalem et~al.}(2017)\citenamefont{Almoalem, Yagil,
  Cho, Teknowijoyo, Tanatar, Prozorov, Liu, Lograsso, and
  Auslaender}}]{Auslaender-2}
\bibinfo{author}{\bibfnamefont{A.}~\bibnamefont{Almoalem}},
  \bibinfo{author}{\bibfnamefont{A.}~\bibnamefont{Yagil}},
  \bibinfo{author}{\bibfnamefont{K.}~\bibnamefont{Cho}},
  \bibinfo{author}{\bibfnamefont{S.}~\bibnamefont{Teknowijoyo}},
  \bibinfo{author}{\bibfnamefont{M.}~\bibnamefont{Tanatar}},
  \bibinfo{author}{\bibfnamefont{R.}~\bibnamefont{Prozorov}},
  \bibinfo{author}{\bibfnamefont{Y.}~\bibnamefont{Liu}},
  \bibinfo{author}{\bibfnamefont{T.}~\bibnamefont{Lograsso}}, \bibnamefont{and}
  \bibinfo{author}{\bibfnamefont{O.}~\bibnamefont{Auslaender}},
  \bibinfo{journal}{preprint arXiv:1708.00683}  (\bibinfo{year}{2017}).

\bibitem[{\citenamefont{Levchenko et~al.}(2013)\citenamefont{Levchenko,
  Vavilov, Khodas, and Chubukov}}]{Levchenko}
\bibinfo{author}{\bibfnamefont{A.}~\bibnamefont{Levchenko}},
  \bibinfo{author}{\bibfnamefont{M.~G.} \bibnamefont{Vavilov}},
  \bibinfo{author}{\bibfnamefont{M.}~\bibnamefont{Khodas}}, \bibnamefont{and}
  \bibinfo{author}{\bibfnamefont{A.~V.} \bibnamefont{Chubukov}},
  \bibinfo{journal}{Phys. Rev. Lett.} \textbf{\bibinfo{volume}{110}},
  \bibinfo{pages}{177003} (\bibinfo{year}{2013}).

\bibitem[{\citenamefont{Chowdhury et~al.}(2013)\citenamefont{Chowdhury,
  Swingle, Berg, and Sachdev}}]{Chowdhury-1}
\bibinfo{author}{\bibfnamefont{D.}~\bibnamefont{Chowdhury}},
  \bibinfo{author}{\bibfnamefont{B.}~\bibnamefont{Swingle}},
  \bibinfo{author}{\bibfnamefont{E.}~\bibnamefont{Berg}}, \bibnamefont{and}
  \bibinfo{author}{\bibfnamefont{S.}~\bibnamefont{Sachdev}},
  \bibinfo{journal}{Phys. Rev. Lett.} \textbf{\bibinfo{volume}{111}},
  \bibinfo{pages}{157004} (\bibinfo{year}{2013}).

\bibitem[{\citenamefont{Nomoto and Ikeda}(2013)}]{Nomoto}
\bibinfo{author}{\bibfnamefont{T.}~\bibnamefont{Nomoto}} \bibnamefont{and}
  \bibinfo{author}{\bibfnamefont{H.}~\bibnamefont{Ikeda}},
  \bibinfo{journal}{Phys. Rev. Lett.} \textbf{\bibinfo{volume}{111}},
  \bibinfo{pages}{167001} (\bibinfo{year}{2013}).

\bibitem[{\citenamefont{Chowdhury et~al.}(2015)\citenamefont{Chowdhury,
  Orenstein, Sachdev, and Senthil}}]{Chowdhury-2}
\bibinfo{author}{\bibfnamefont{D.}~\bibnamefont{Chowdhury}},
  \bibinfo{author}{\bibfnamefont{J.}~\bibnamefont{Orenstein}},
  \bibinfo{author}{\bibfnamefont{S.}~\bibnamefont{Sachdev}}, \bibnamefont{and}
  \bibinfo{author}{\bibfnamefont{T.}~\bibnamefont{Senthil}},
  \bibinfo{journal}{Phys. Rev. B} \textbf{\bibinfo{volume}{92}},
  \bibinfo{pages}{081113} (\bibinfo{year}{2015}).

\bibitem[{\citenamefont{Walmsley et~al.}(2013)\citenamefont{Walmsley, Putzke,
  Malone, Guillam\'on, Vignolles, Proust, Badoux, Coldea, Watson, Kasahara
  et~al.}}]{Carrington-1}
\bibinfo{author}{\bibfnamefont{P.}~\bibnamefont{Walmsley}},
  \bibinfo{author}{\bibfnamefont{C.}~\bibnamefont{Putzke}},
  \bibinfo{author}{\bibfnamefont{L.}~\bibnamefont{Malone}},
  \bibinfo{author}{\bibfnamefont{I.}~\bibnamefont{Guillam\'on}},
  \bibinfo{author}{\bibfnamefont{D.}~\bibnamefont{Vignolles}},
  \bibinfo{author}{\bibfnamefont{C.}~\bibnamefont{Proust}},
  \bibinfo{author}{\bibfnamefont{S.}~\bibnamefont{Badoux}},
  \bibinfo{author}{\bibfnamefont{A.~I.} \bibnamefont{Coldea}},
  \bibinfo{author}{\bibfnamefont{M.~D.} \bibnamefont{Watson}},
  \bibinfo{author}{\bibfnamefont{S.}~\bibnamefont{Kasahara}},
  \bibnamefont{et~al.}, \bibinfo{journal}{Phys. Rev. Lett.}
  \textbf{\bibinfo{volume}{110}}, \bibinfo{pages}{257002}
  (\bibinfo{year}{2013}).

\bibitem[{\citenamefont{Putzke et~al.}(2014)\citenamefont{Putzke, Walmsley,
  Fletcher, Malone, Vignolles, Proust, Badoux, See, Beere, Ritchie
  et~al.}}]{Carrington-2}
\bibinfo{author}{\bibfnamefont{C.}~\bibnamefont{Putzke}},
  \bibinfo{author}{\bibfnamefont{P.}~\bibnamefont{Walmsley}},
  \bibinfo{author}{\bibfnamefont{J.~D.} \bibnamefont{Fletcher}},
  \bibinfo{author}{\bibfnamefont{L.}~\bibnamefont{Malone}},
  \bibinfo{author}{\bibfnamefont{D.}~\bibnamefont{Vignolles}},
  \bibinfo{author}{\bibfnamefont{C.}~\bibnamefont{Proust}},
  \bibinfo{author}{\bibfnamefont{S.}~\bibnamefont{Badoux}},
  \bibinfo{author}{\bibfnamefont{P.}~\bibnamefont{See}},
  \bibinfo{author}{\bibfnamefont{H.~E.} \bibnamefont{Beere}},
  \bibinfo{author}{\bibfnamefont{D.~A.} \bibnamefont{Ritchie}},
  \bibnamefont{et~al.}, \bibinfo{journal}{Nature Communications}
  \textbf{\bibinfo{volume}{5}}, \bibinfo{pages}{5679} (\bibinfo{year}{2014}).

\bibitem[{\citenamefont{Parker and Mazin}(2009)}]{Mazin}
\bibinfo{author}{\bibfnamefont{D.}~\bibnamefont{Parker}} \bibnamefont{and}
  \bibinfo{author}{\bibfnamefont{I.~I.} \bibnamefont{Mazin}},
  \bibinfo{journal}{Phys. Rev. Lett.} \textbf{\bibinfo{volume}{102}},
  \bibinfo{pages}{227007} (\bibinfo{year}{2009}).

\bibitem[{\citenamefont{Wu and Phillips}(2009)}]{Wu}
\bibinfo{author}{\bibfnamefont{J.}~\bibnamefont{Wu}} \bibnamefont{and}
  \bibinfo{author}{\bibfnamefont{P.}~\bibnamefont{Phillips}},
  \bibinfo{journal}{Phys. Rev. B} \textbf{\bibinfo{volume}{79}},
  \bibinfo{pages}{092502} (\bibinfo{year}{2009}).

\bibitem[{\citenamefont{Ng and Nagaosa}(2009)}]{Nagaosa}
\bibinfo{author}{\bibfnamefont{T.~K.} \bibnamefont{Ng}} \bibnamefont{and}
  \bibinfo{author}{\bibfnamefont{N.}~\bibnamefont{Nagaosa}},
  \bibinfo{journal}{EPL (Europhysics Letters)} \textbf{\bibinfo{volume}{87}},
  \bibinfo{pages}{17003} (\bibinfo{year}{2009}).

\bibitem[{\citenamefont{Linder et~al.}(2009)\citenamefont{Linder, Sperstad, and
  Sudb\o{}}}]{Linder}
\bibinfo{author}{\bibfnamefont{J.}~\bibnamefont{Linder}},
  \bibinfo{author}{\bibfnamefont{I.~B.} \bibnamefont{Sperstad}},
  \bibnamefont{and} \bibinfo{author}{\bibfnamefont{A.}~\bibnamefont{Sudb\o{}}},
  \bibinfo{journal}{Phys. Rev. B} \textbf{\bibinfo{volume}{80}},
  \bibinfo{pages}{020503} (\bibinfo{year}{2009}).

\bibitem[{\citenamefont{Tsai et~al.}(2009)\citenamefont{Tsai, Yao, Bernevig,
  and Hu}}]{Tsai}
\bibinfo{author}{\bibfnamefont{W.-F.} \bibnamefont{Tsai}},
  \bibinfo{author}{\bibfnamefont{D.-X.} \bibnamefont{Yao}},
  \bibinfo{author}{\bibfnamefont{B.~A.} \bibnamefont{Bernevig}},
  \bibnamefont{and} \bibinfo{author}{\bibfnamefont{J.}~\bibnamefont{Hu}},
  \bibinfo{journal}{Phys. Rev. B} \textbf{\bibinfo{volume}{80}},
  \bibinfo{pages}{012511} (\bibinfo{year}{2009}).

\bibitem[{\citenamefont{Chen et~al.}(2009)\citenamefont{Chen, Ma, Lu, and
  Zhang}}]{Chen}
\bibinfo{author}{\bibfnamefont{W.-Q.} \bibnamefont{Chen}},
  \bibinfo{author}{\bibfnamefont{F.}~\bibnamefont{Ma}},
  \bibinfo{author}{\bibfnamefont{Z.-Y.} \bibnamefont{Lu}}, \bibnamefont{and}
  \bibinfo{author}{\bibfnamefont{F.-C.} \bibnamefont{Zhang}},
  \bibinfo{journal}{Phys. Rev. Lett.} \textbf{\bibinfo{volume}{103}},
  \bibinfo{pages}{207001} (\bibinfo{year}{2009}).

\bibitem[{\citenamefont{Ota et~al.}(2009)\citenamefont{Ota, Machida, Koyama,
  and Matsumoto}}]{Ota-1}
\bibinfo{author}{\bibfnamefont{Y.}~\bibnamefont{Ota}},
  \bibinfo{author}{\bibfnamefont{M.}~\bibnamefont{Machida}},
  \bibinfo{author}{\bibfnamefont{T.}~\bibnamefont{Koyama}}, \bibnamefont{and}
  \bibinfo{author}{\bibfnamefont{H.}~\bibnamefont{Matsumoto}},
  \bibinfo{journal}{Phys. Rev. Lett.} \textbf{\bibinfo{volume}{102}},
  \bibinfo{pages}{237003} (\bibinfo{year}{2009}).

\bibitem[{\citenamefont{Ota et~al.}(2010)\citenamefont{Ota, Nakai, Nakamura,
  Machida, Inotani, Ohashi, Koyama, and Matsumoto}}]{Ota-2}
\bibinfo{author}{\bibfnamefont{Y.}~\bibnamefont{Ota}},
  \bibinfo{author}{\bibfnamefont{N.}~\bibnamefont{Nakai}},
  \bibinfo{author}{\bibfnamefont{H.}~\bibnamefont{Nakamura}},
  \bibinfo{author}{\bibfnamefont{M.}~\bibnamefont{Machida}},
  \bibinfo{author}{\bibfnamefont{D.}~\bibnamefont{Inotani}},
  \bibinfo{author}{\bibfnamefont{Y.}~\bibnamefont{Ohashi}},
  \bibinfo{author}{\bibfnamefont{T.}~\bibnamefont{Koyama}}, \bibnamefont{and}
  \bibinfo{author}{\bibfnamefont{H.}~\bibnamefont{Matsumoto}},
  \bibinfo{journal}{Phys. Rev. B} \textbf{\bibinfo{volume}{81}},
  \bibinfo{pages}{214511} (\bibinfo{year}{2010}).

\bibitem[{\citenamefont{Yerin and Omelyanchouk}(2010)}]{Yerin}
\bibinfo{author}{\bibfnamefont{Y.~S.} \bibnamefont{Yerin}} \bibnamefont{and}
  \bibinfo{author}{\bibfnamefont{A.~N.} \bibnamefont{Omelyanchouk}},
  \bibinfo{journal}{Low Temperature Physics} \textbf{\bibinfo{volume}{36}},
  \bibinfo{pages}{969} (\bibinfo{year}{2010}).

\bibitem[{\citenamefont{Berg et~al.}(2011)\citenamefont{Berg, Lindner, and
  Pereg-Barnea}}]{Berg}
\bibinfo{author}{\bibfnamefont{E.}~\bibnamefont{Berg}},
  \bibinfo{author}{\bibfnamefont{N.~H.} \bibnamefont{Lindner}},
  \bibnamefont{and}
  \bibinfo{author}{\bibfnamefont{T.}~\bibnamefont{Pereg-Barnea}},
  \bibinfo{journal}{Phys. Rev. Lett.} \textbf{\bibinfo{volume}{106}},
  \bibinfo{pages}{147003} (\bibinfo{year}{2011}).

\bibitem[{\citenamefont{Golubov and Mazin}(2013)}]{Golubov}
\bibinfo{author}{\bibfnamefont{A.~A.} \bibnamefont{Golubov}} \bibnamefont{and}
  \bibinfo{author}{\bibfnamefont{I.~I.} \bibnamefont{Mazin}},
  \bibinfo{journal}{Applied Physics Letters} \textbf{\bibinfo{volume}{102}},
  \bibinfo{pages}{032601} (\bibinfo{year}{2013}).

\bibitem[{\citenamefont{Koshelev and Stanev}(2011)}]{Koshelev}
\bibinfo{author}{\bibfnamefont{A.~E.} \bibnamefont{Koshelev}} \bibnamefont{and}
  \bibinfo{author}{\bibfnamefont{V.}~\bibnamefont{Stanev}},
  \bibinfo{journal}{EPL (Europhysics Letters)} \textbf{\bibinfo{volume}{96}},
  \bibinfo{pages}{27014} (\bibinfo{year}{2011}).

\bibitem[{\citenamefont{Stanev and Koshelev}(2012)}]{Stanev}
\bibinfo{author}{\bibfnamefont{V.~G.} \bibnamefont{Stanev}} \bibnamefont{and}
  \bibinfo{author}{\bibfnamefont{A.~E.} \bibnamefont{Koshelev}},
  \bibinfo{journal}{Phys. Rev. B} \textbf{\bibinfo{volume}{86}},
  \bibinfo{pages}{174515} (\bibinfo{year}{2012}).

\bibitem[{\citenamefont{Lin}(2012)}]{Lin}
\bibinfo{author}{\bibfnamefont{S.-Z.} \bibnamefont{Lin}},
  \bibinfo{journal}{Phys. Rev. B} \textbf{\bibinfo{volume}{86}},
  \bibinfo{pages}{014510} (\bibinfo{year}{2012}).

\bibitem[{\citenamefont{Vakaryuk et~al.}(2012)\citenamefont{Vakaryuk, Stanev,
  Lee, and Levchenko}}]{Vakaryuk}
\bibinfo{author}{\bibfnamefont{V.}~\bibnamefont{Vakaryuk}},
  \bibinfo{author}{\bibfnamefont{V.}~\bibnamefont{Stanev}},
  \bibinfo{author}{\bibfnamefont{W.-C.} \bibnamefont{Lee}}, \bibnamefont{and}
  \bibinfo{author}{\bibfnamefont{A.}~\bibnamefont{Levchenko}},
  \bibinfo{journal}{Phys. Rev. Lett.} \textbf{\bibinfo{volume}{109}},
  \bibinfo{pages}{227003} (\bibinfo{year}{2012}).

\bibitem[{\citenamefont{Apostolov and Levchenko}(2012)}]{Apostolov}
\bibinfo{author}{\bibfnamefont{S.}~\bibnamefont{Apostolov}} \bibnamefont{and}
  \bibinfo{author}{\bibfnamefont{A.}~\bibnamefont{Levchenko}},
  \bibinfo{journal}{Phys. Rev. B} \textbf{\bibinfo{volume}{86}},
  \bibinfo{pages}{224501} (\bibinfo{year}{2012}).

\bibitem[{\citenamefont{Vavilov and Chubukov}(2011)}]{Vavilov}
\bibinfo{author}{\bibfnamefont{M.~G.} \bibnamefont{Vavilov}} \bibnamefont{and}
  \bibinfo{author}{\bibfnamefont{A.~V.} \bibnamefont{Chubukov}},
  \bibinfo{journal}{Phys. Rev. B} \textbf{\bibinfo{volume}{84}},
  \bibinfo{pages}{214521} (\bibinfo{year}{2011}).

\bibitem[{\citenamefont{Moor et~al.}(2011)\citenamefont{Moor, Volkov, and
  Efetov}}]{Moor}
\bibinfo{author}{\bibfnamefont{A.}~\bibnamefont{Moor}},
  \bibinfo{author}{\bibfnamefont{A.~F.} \bibnamefont{Volkov}},
  \bibnamefont{and} \bibinfo{author}{\bibfnamefont{K.~B.}
  \bibnamefont{Efetov}}, \bibinfo{journal}{Phys. Rev. B}
  \textbf{\bibinfo{volume}{83}}, \bibinfo{pages}{134524}
  (\bibinfo{year}{2011}).

\bibitem[{\citenamefont{Dzero et~al.}(2015)\citenamefont{Dzero, Khodas,
  Klironomos, Vavilov, and Levchenko}}]{Dzero}
\bibinfo{author}{\bibfnamefont{M.}~\bibnamefont{Dzero}},
  \bibinfo{author}{\bibfnamefont{M.}~\bibnamefont{Khodas}},
  \bibinfo{author}{\bibfnamefont{A.~D.} \bibnamefont{Klironomos}},
  \bibinfo{author}{\bibfnamefont{M.~G.} \bibnamefont{Vavilov}},
  \bibnamefont{and}
  \bibinfo{author}{\bibfnamefont{A.}~\bibnamefont{Levchenko}},
  \bibinfo{journal}{Phys. Rev. B} \textbf{\bibinfo{volume}{92}},
  \bibinfo{pages}{144501} (\bibinfo{year}{2015}).


\bibitem[{\citenamefont{Fernandes et~al.}(2012)\citenamefont{Fernandes,
  Vavilov, and Chubukov}}]{Fernandes-Tc}
\bibinfo{author}{\bibfnamefont{R.~M.} \bibnamefont{Fernandes}},
  \bibinfo{author}{\bibfnamefont{M.~G.} \bibnamefont{Vavilov}},
  \bibnamefont{and} \bibinfo{author}{\bibfnamefont{A.~V.}
  \bibnamefont{Chubukov}}, \bibinfo{journal}{Phys. Rev. B}
  \textbf{\bibinfo{volume}{85}}, \bibinfo{pages}{140512}
  (\bibinfo{year}{2012}).

\bibitem[{\citenamefont{Zaitsev}(1984)}]{Zaitsev1984}
\bibinfo{author}{\bibfnamefont{A.~V.} \bibnamefont{Zaitsev}},
  \bibinfo{journal}{Sov. Phys. JETP} \textbf{\bibinfo{volume}{59}},
  \bibinfo{pages}{1015} (\bibinfo{year}{1984}).

\bibitem[{\citenamefont{Yip}(1997)}]{Yip1997}
\bibinfo{author}{\bibfnamefont{S.-K.} \bibnamefont{Yip}},
  \bibinfo{journal}{Jour. of Low Temp. Phys.} \textbf{\bibinfo{volume}{109}},
  \bibinfo{pages}{547} (\bibinfo{year}{1997}).

\bibitem[{\citenamefont{Nazarov}(1999)}]{Nazarov}
\bibinfo{author}{\bibfnamefont{Y.~V.} \bibnamefont{Nazarov}},
  \bibinfo{journal}{Superlattices and Microstructures}
  \textbf{\bibinfo{volume}{25}}, \bibinfo{pages}{1221 } (\bibinfo{year}{1999}).

\bibitem[{\citenamefont{Baranger and Mello}(1994)}]{Baranger}
\bibinfo{author}{\bibfnamefont{H.~U.} \bibnamefont{Baranger}} \bibnamefont{and}
  \bibinfo{author}{\bibfnamefont{P.~A.} \bibnamefont{Mello}},
  \bibinfo{journal}{Phys. Rev. Lett.} \textbf{\bibinfo{volume}{73}},
  \bibinfo{pages}{142} (\bibinfo{year}{1994}).

\bibitem[{\citenamefont{Dorokhov}(1984)}]{Dorokhov}
\bibinfo{author}{\bibfnamefont{O.}~\bibnamefont{Dorokhov}},
  \bibinfo{journal}{Solid State Communications} \textbf{\bibinfo{volume}{51}},
  \bibinfo{pages}{381 } (\bibinfo{year}{1984}).

\bibitem[{\citenamefont{Schep and Bauer}(1997)}]{Schep}
\bibinfo{author}{\bibfnamefont{K.~M.} \bibnamefont{Schep}} \bibnamefont{and}
  \bibinfo{author}{\bibfnamefont{G.~E.~W.} \bibnamefont{Bauer}},
  \bibinfo{journal}{Phys. Rev. Lett.} \textbf{\bibinfo{volume}{78}},
  \bibinfo{pages}{3015} (\bibinfo{year}{1997}).

\bibitem{Ishida}
Shigeyuki Ishida, Dongjoon Song, Hiraku Ogino, Akira Iyo, and Hiroshi Eisaki, Masamichi Nakajima, 
Jun-ichi Shimoyama, and Michael Eisterer, Phys. Rev. B \textbf{95}, 014517 (2017). 
 
\bibitem[{\citenamefont{Mizukami et~al.}(2014)\citenamefont{Mizukami,
  Konczykowski, Kawamoto, Kurata, Kasahara, Hashimoto, Mishra, Kreisel, Wang,
  Hirschfeld et~al.}}]{Mizukami}
\bibinfo{author}{\bibfnamefont{Y.}~\bibnamefont{Mizukami}},
  \bibinfo{author}{\bibfnamefont{M.}~\bibnamefont{Konczykowski}},
  \bibinfo{author}{\bibfnamefont{Y.}~\bibnamefont{Kawamoto}},
  \bibinfo{author}{\bibfnamefont{S.}~\bibnamefont{Kurata}},
  \bibinfo{author}{\bibfnamefont{S.}~\bibnamefont{Kasahara}},
  \bibinfo{author}{\bibfnamefont{K.}~\bibnamefont{Hashimoto}},
  \bibinfo{author}{\bibfnamefont{V.}~\bibnamefont{Mishra}},
  \bibinfo{author}{\bibfnamefont{A.}~\bibnamefont{Kreisel}},
  \bibinfo{author}{\bibfnamefont{Y.}~\bibnamefont{Wang}},
  \bibinfo{author}{\bibfnamefont{P.~J.} \bibnamefont{Hirschfeld}},
  \bibnamefont{et~al.}, \bibinfo{journal}{Nature Communications}
  \textbf{\bibinfo{volume}{5}}, \bibinfo{pages}{5657} (\bibinfo{year}{2014}).

\bibitem[{\citenamefont{Jung et~al.}(2018)\citenamefont{Jung, Seo, Lee, Bauer,
  Lee, and Park}}]{Bauer2018}
\bibinfo{author}{\bibfnamefont{S.-G.} \bibnamefont{Jung}},
  \bibinfo{author}{\bibfnamefont{S.}~\bibnamefont{Seo}},
  \bibinfo{author}{\bibfnamefont{S.}~\bibnamefont{Lee}},
  \bibinfo{author}{\bibfnamefont{E.~D.} \bibnamefont{Bauer}},
  \bibinfo{author}{\bibfnamefont{H.-O.} \bibnamefont{Lee}}, \bibnamefont{and}
  \bibinfo{author}{\bibfnamefont{T.}~\bibnamefont{Park}},
  \bibinfo{journal}{preprint arXiv:1804.01376}  (\bibinfo{year}{2018}).

\bibitem[{\citenamefont{Talantsev et~al.}(2017)\citenamefont{Talantsev, Crump,
  and Tallon}}]{Talantsev}
\bibinfo{author}{\bibfnamefont{E.}~\bibnamefont{Talantsev}},
  \bibinfo{author}{\bibfnamefont{W.~P.} \bibnamefont{Crump}}, \bibnamefont{and}
  \bibinfo{author}{\bibfnamefont{J.~L.} \bibnamefont{Tallon}},
  \bibinfo{journal}{Annalen der Physik} \textbf{\bibinfo{volume}{529}},
  \bibinfo{pages}{1700197} (\bibinfo{year}{2017}).

\bibitem[{\citenamefont{Abrikosov}(1988)}]{TheoryOfMetals}
\bibinfo{author}{\bibfnamefont{A.~A.} \bibnamefont{Abrikosov}},
  \emph{\bibinfo{title}{Fundamentals of the Theory of Metals}}
  (\bibinfo{publisher}{North-Holland, Elsevier Science Publishers},
  \bibinfo{year}{1988}).

\end{thebibliography}
\end{document}